\documentclass[preprint,showpacs,preprintnumbers,amsmath,amssymb,superscriptaddress]{revtex4}

\usepackage{graphicx}% Include figure files
\usepackage{bm}% bold math

\usepackage{color,graphics}
\definecolor{orange}{rgb}{.95,.75,0}

\begin{document}
\def\mstar{m_{\displaystyle *}}
\def\revise{\color{red}}

\title{Breather mode in the many-electron dynamics of semiconductor quantum wells}
\author{F. Haas}
\affiliation{Institut f\"ur Theoretische Physik IV,
Ruhr-Universit\"at Bochum, D-44780 Bochum, Germany}
\author{G. Manfredi} \affiliation{Institut de Physique et Chimie des
Mat\'{e}riaux de Strasbourg, BP 43, F-67034 Strasbourg, France}
\author{P. K. Shukla}
\affiliation{Institut f\"ur Theoretische Physik IV,
Ruhr-Universit\"at Bochum, D-44780 Bochum, Germany}
\author{P.-A.Hervieux} \affiliation{Institut de Physique et Chimie des
Mat\'{e}riaux de Strasbourg, BP 43, F-67034 Strasbourg, France}

%\date{\today}

\begin{abstract}
\noindent We demonstrate the existence of a novel breather mode in
the self-consistent electron dynamics of a semiconductor quantum
well. A non-perturbative variational method based on quantum
hydrodynamics is used to determine the salient features of the
electron breather mode. Numerical simulations of the
time-dependent Wigner-Poisson or Hartree equations are shown to be
in excellent agreement with our analytical results. For
asymmetric quantum wells, a signature of the breather mode is
observed in the dipole response, which can be detected by standard
optical means.
\end{abstract}


\maketitle

{\it Introduction}.--- The many-electron dynamics in nanoscale
semiconductor devices, such as quantum wells and quantum dots, has
recently attracted a great deal of interest, mainly in view of
possible applications to the growing field of quantum computing
\cite{Zoller}. Particular attention has been devoted to
intersubband transitions, which involve excitation frequencies of
the order of the terahertz \cite{Heyman}. On this time scale,
various collective electronic modes can be excited. For instance,
the electric dipole response is dominated by a strong resonance at
the effective plasmon frequency. This resonance (known as the Kohn
mode \cite{Kohn}) is characterized by rigid oscillations of the
electron gas, which, for perfectly parabolic confinement, are
decoupled from the internal degrees of freedom.

In this paper, we show the existence of a new distinct resonance
-- a monopole or ``breather'' mode -- which corresponds to
coherent oscillations of the size of the electron gas around a
self-consistent equilibrium. Breather modes have been described in
many areas of physics, such as nuclear matter \cite{Harakesh,
Youngblood} (where they are known as giant monopole resonances)
and ultracold atom dynamics \cite{Goral,Zhou}. In experiments on
metallic nanoparticles, monopole oscillations of the {\em ionic}
structure have been observed, which manifest themselves as slow
modulations of the surface plasmon \cite{Portales}. However, to
the best of our knowledge, previous investigations have not
addressed the features of the breather mode in the self-consistent
dynamics of a confined electron gas. Although quantum wells
constitute a typical instance of such confined systems, the
present approach should equally apply to metal nanoparticles and
carbon-based systems such as fullerenes.

{\it Model}.---Because of the translational symmetry in
the transverse plane, the problem reduces to a one-dimensional
(1D) one in the $x$ direction \cite{Santer,Wijewardane}. To model
the electron dynamics, we use a self-consistent quantum
hydrodynamic model (QHM) that was originally derived for quantum
plasmas \cite{Manfredi,Manfredi2} and metallic nanostructures
\cite{Crouseilles}. In the QHM, the evolution of the electron
density $n(x,t)$ and mean velocity $u(x,t)$ is governed by the
continuity and momentum equations
%1
\begin{eqnarray}
\label{e1}
\frac{\partial n}{\partial t} &+& \frac{\partial}{\partial x}\,(nu) = 0 \,,\\
 \frac{\partial\,u}{\partial\,t} &+& u \frac{\partial
u}{\partial x} = -\frac{1}{\mstar n} \frac{\partial P}{\partial x}
\label{e2} - \frac{1}\mstar\frac{\partial V_{\rm eff}}{\partial x}
+ \frac{\hbar^2}{2\mstar^2} ~\frac{\partial}{\partial x}
\left(\frac{\partial_x^2 \sqrt{n}}{\sqrt{n}} \right) ,
\end{eqnarray}
where $\mstar$ is the effective electron mass, $\hbar$ is the
reduced Planck constant, $P(x,t)$ is the electron pressure, and
$V_{\rm eff}=V_{\rm conf}(x)+V_H(x,t)$ is the effective potential,
which is composed of a confining and a Hartree term. The Hartree
potential obeys the Poisson equation, namely $V_H''= -
e^{2}\,n/\varepsilon$, where $e$ is the magnitude of the electron
charge and $\varepsilon$ is the effective dielectric permeability
of the material.  The term proportional to $\hbar^2$ on the
right-hand side of Eq. (\ref{e2}) represents the quantum force due
to the so-called Bohm potential \cite{Gardner}.

The above QHM can be derived from the self-consistent Hartree
equations \cite{Manfredi2} -- or equivalently from the phase-space
Wigner-Poisson equations \cite{Manfredi3} -- in the limit of long
wavelengths compared to the Thomas-Fermi screening length. For the
sake of simplicity, we shall neglect exchange/correlation
corrections and assume Boltzmann statistics, which is a reasonable
approximation at moderate electron temperatures $T$
\cite{Santer,Gusev}. We also stress that the 1D model
relies on the separation of the transverse and longitudinal
directions, which may be broken by collisional effects. However,
such effects should not be dominant on the fast time scales
considered here \cite{Heyman95}.

The pressure $P(x,t)$ in Eq. (\ref{e2}) must be related to the
electron density $n$ via an equation of state (EOS) in order to
close our system of electron fluid equations. We take a polytropic
relation $P = \overline{n}\,k_B\,T\,(n/\overline{n})^{\gamma}$,
where $k_B$ is the Boltzmann constant, $\gamma=3$ is the 1D
polytropic exponent, and $\overline{n}$ is a mean electron
density. For a homogeneous system (where $\overline{n}=n_0$), this
EOS correctly reproduces the Bohm-Gross dispersion relation
\cite{Manfredi2} in quantum plasmas. For the inhomogeneous
electron gas considered here, the choice of $\overline{n}$ is
subtler and will be discussed later.

We assume a parabolic confinement, with $V_{\rm conf} =
\frac{1}{2}\omega_0^2 \mstar x^2$, where the frequency $\omega_0$
can be related to a fictitious homogeneous positive charge of
density $n_0$ via the relation $\omega_0 = (e^2 n_0/\mstar
\epsilon)^{1/2} $. We then normalize time to $\omega_0^{-1}$;
space to $L_{0} = (k_{B}\,T/\mstar)^{1/2}/\omega_{0}$; velocity to
$L_{0}\,\omega_0$; energy to $k_B T$; and the electron number
density to $n_0$. Quantum effects are measured by the
dimensionless parameter $H = \hbar\,\omega_{0}/k_{B}\,T$.

We shall use typical parameters that are representative of
semiconductor quantum wells \cite{Wijewardane}: the effective
electron mass and the effective dielectric permeability are,
respectively, $\mstar=0.067m_e$ and $\varepsilon =
13\,\varepsilon_0$, the equilibrium density is $n_0 = 4.7 \times
10^{22}\,{\rm m}^{-3}$, and the filling fraction
$\overline{n}/n_0=0.5$. These values yield an effective plasmon
energy $\hbar\,\omega_0 = 8.62 \,{\rm meV}$, a characteristic
length $L_0 = 16.2 \,{\rm nm}$, a Fermi temperature $T_F=51.8$ K,
and a typical time scale $\omega_0^{-1}=76$ fs. An electron
temperature $T = 200\,{\rm K}$ then corresponds to a value $H =
0.5$.

{\it Lagrangian approach}.--- In order to derive a closed system
of differential equations describing the breather mode, we fist
express the quantum hydrodynamical equations in a Lagrangian
formalism. We stress that this approach is not based on a
perturbative expansion, and thus is not restricted to the linear
regime. The Lagrangian density corresponding to the system of Eqs.
(\ref{e1})--(\ref{e2}) reads as (normalized units are used from
now on)
%2
\begin{eqnarray}
{\cal L} &=& \frac{1}{2}\left(\frac{\partial V_H}{\partial x}
\right)^2 - n\,V_H - \,n\,\frac{\partial\theta}{\partial t} -
\int^{n}W(n')\,dn' \nonumber \\ \label{e4} &-&
\frac{1}{2}\,\left(n\,\left[\frac{\partial\theta}{\partial
x}\right]^2 + \frac{H^2}{4n}\,\left[\frac{\partial\,n} {\partial
x}\right]^2\right) - n\,V_{\rm conf} \,,
\end{eqnarray}
where the independent fields are taken to be $n$, $\theta$, and
$V_H$. The velocity field follows from the auxiliary function
$\theta(x,t)$ through $u = \partial\theta/\partial x$. The
quantity $W(n)$ in Eq. (\ref{e4}) originates from the pressure, $W
\equiv \int^{n} \frac{dP}{dn'}\frac{dn'}{n'} =
(3/2)(n/\overline{n})^2$. Taking the variational derivatives of
the action $S = \int\,{\cal L}\,dx\,dt$ with respect to $n$,
$\theta$, and $V_H$, we obtain the Eqs. (\ref{e1})--(\ref{e2}), as
well as the Poisson equation for $V_H$.

The existence of a pertinent variational formalism can be used to
derive approximate solutions via the time-dependent Rayleigh-Ritz
trial-function method \cite{Matuszewski}. For this purpose, we
assume the electron density to have a Gaussian profile
\begin{equation}
\label{e5} n(x,t) = \frac{A}{\sigma}\,\exp\left[-
\frac{(x-d\,)^2}{2\,\sigma^2}\right] \,,
\end{equation}
where $d(t)$ and $\sigma(t)$ are time-dependent functions that
represent the center-of-mass (dipole) and the spatial dispersion
of the electron gas, respectively. The constant $A =
N/\sqrt{2\pi}$, is related to the total number of electrons in the
well, $N=\int n\,dx$. The above {\it Ansatz} is a natural one,
because for a negligible Hartree energy, $V_{\rm eff}$ reduces to
a harmonic oscillator potential.

The other fields to be inserted in the action functional are
$\theta$ and $V_H$. The natural way to choose them is by requiring
that the continuity and Poisson equations are automatically
satisfied. The continuity equation is solved with $n$ given by Eq.
(\ref{e5}) together with $u = \dot{d} + (\dot\sigma/\sigma)\,\xi$,
which leads to $\theta = (\dot\sigma/2\sigma)\,\xi^2 +
\dot{d}\,\xi$, where $\xi\equiv x-d$.  An irrelevant gauge
function was discarded in the calculation of $\theta$. The
solution of the Poisson equation with a Gaussian electron density
is
\begin{equation}
\label{e6} V_H = - A\,\sigma\,e^{-\xi^2/2\sigma^2}
 - A\,\sqrt{\frac{\pi}{2}}\,\xi\,{\rm
Erf}\left(\frac{\xi}{\sqrt{2}\,\sigma}\right) + \rm const,
\end{equation}
where ${\rm Erf}$ is the error function. The integration constant
is chosen so that $V_H(\pm a)=0$, with $2a$ being the total size
of the system, and letting $a\to\infty$ at the end of the
calculation. As the potential $V_H$ is not bounded, a divergence
appears in the Lagrangian density when integrated over space.
However, the divergent term does not depend on the dynamical
variables $d$ and $\sigma$, so that it can be ignored.
% The divergence is due to the peculiar properties of the electric
% potential in 1D, and disappears in 3D.

Using the above {\it Ansatz}, one obtains the Lagrangian
\begin{eqnarray}
L &\equiv& \frac{1}{\sqrt{2\pi}\,A}\,\int\,{\cal L}\,dx =
\frac{\dot{d}^2
+ {\dot\sigma}^2}{2} - \frac{d^2 + \sigma^2}{2}  \nonumber \\
&+& \frac{\sqrt{2}}{2}\,A\,\sigma -
\,\frac{\sqrt{3}\,A^2}{6\,\overline{n}^2\,\sigma^2} -
\frac{H^2}{8\,\sigma^2} \,, \label{lag}
\end{eqnarray}
which only depends on two degrees of freedom, namely the dipole
$d$ and the variance $\sigma$. The Euler-Lagrange equations
corresponding to the Lagrangian $L$ read as
\begin{eqnarray}
\label{e7} \ddot{d} &+& d = 0 \\ \label{e8} \ddot{\sigma} &+&
\sigma = \frac{\sqrt{2}\,A}{2} +
\frac{\sqrt{3}\,A^2}{3\,\overline{n}^2\,\sigma^3}  +
\frac{H^2}{4\,\sigma^3} \,.
\end{eqnarray}

\begin{figure}[htb]
\centering
\includegraphics[width=6.5cm,height=5.cm]{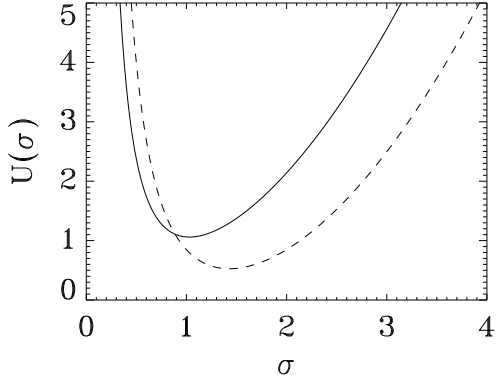}
\caption{Profiles of the pseudo-potential $U(\sigma)$, for $H =
0.5$, $A = 0$ (solid line) and $H = 0.5$, $A = 1$ (dashed line).
The fixed points are $\sigma_0 = 1.03$ ($A = 0$) and $\sigma_0 =
1.43$ ($A = 1$).} \label{fig1}
\end{figure}

The quantum-well potential $V_{\rm conf}$ manifests itself in the
harmonic forces on the left-hand side of both Eqs. (\ref{e7}) and
(\ref{e8}). As expected, the equations for $d$ and $\sigma$
decouple for purely harmonic confinement. Equation (\ref{e7})
describes rigid oscillations of the electron gas at the effective
plasmonic frequency, i.e. the Kohn mode \cite{Kohn,Dobson}.
Equation (\ref{e8}) describes the dynamics of the breather mode,
which features coherent oscillations of the width of the electron
density.  The three terms in the right-hand side of Eq. (\ref{e8})
correspond to the Coulomb repulsion (Hartree term), the electron
pressure, and the quantum Bohm potential, respectively. The
breather equation (\ref{e8}) can be written as $\ddot{\sigma} = -
\,dU/d\sigma$, where $U(\sigma)$ is a pseudo-potential defined as
$U =\sigma^2/2 - \sqrt{2}\,A\,\sigma/2 +
\sqrt{3}\,A^{2}/(6\,\overline{n}^2\,\sigma^2) +
H^2/(8\,\sigma^2)$. From the shape of the pseudo-potential (Fig.
\ref{fig1}), it follows that $\sigma$ will always execute
nonlinear oscillations around a stable fixed point
$\sigma_0(A,H)$, which is a solution of the algebraic equation
$U'(\sigma_0)=0$.

So far, we have not specified the value of the average density
$\overline{n}$ that appears in the EOS, $P=n^3/\overline{n}^2$,
written in normalized units. It is natural to assume that
$\overline{n}$ takes some value smaller than the peak density at
equilibrium $A/\sigma_0$. The correct way to compute this value is
to average the square of the density using $n$ itself, i.e.
$\overline{n}^2 \equiv \langle n^2 \rangle = \int n^3 dx/\int ndx
= A^2/(\sqrt{3}\sigma_0^2)$. A useful check can be performed by
plugging this expression into Eq. (\ref{e8}) and neglecting the
Hartree potential, which yields the equilibrium variance $\sigma_0
= [1+(1+H^2)^{1/2}]^{1/2}/\sqrt{2}$. This expression displays the
correct low- and high-temperature limits for the quantum harmonic
oscillator: $\sigma_0 \to 1$, for $H \to 0$; and $\sigma_0 \sim
\sqrt{H/2}$, for $H\gg 1$.

\begin{figure}[htb]
\centering
\includegraphics[width=7.cm]{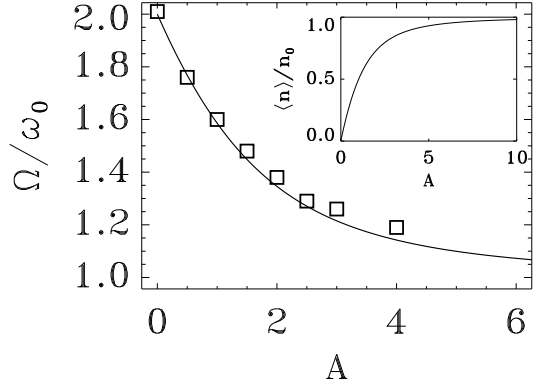}
\caption{The breather frequency $\Omega$ as a function of $A$, for
$H=0.5$. Solid line: analytical results from the Lagrangian
method. Squares: The Wigner-Poisson (WP) simulations. The inset shows
the mean electron density $\langle n\rangle = A/\sqrt{2}\sigma_0$
as a function of $A$.} \label{fig2}
\end{figure}

With the above prescription for $\overline{n}$, the
pseudo-potential becomes $U= \sigma^2/2  - \sqrt{2}\,A\,\sigma/2 +
\sigma_{0}^2/(2\,\sigma^2) + H^2/(8\,\sigma^2)$. The frequency
$\Omega=\Omega(A,H)$ of the breather mode, corresponding to the
oscillations of $\sigma$, can be obtained by linearizing the
equation of motion [Eq. (\ref{e8})] in the vicinity of the stable
fixed point of $U(\sigma)$. The dependence of the breather
frequency with $A$ (i.e., with the electron density) is shown in
Fig. \ref{fig2}. For $A=0$ (i.e., without the Coulomb interaction)
the exact frequency is $\Omega=2\omega_0$. For finite $A$, the
breather frequency decreases and approaches $\Omega=\omega_0$, for
$A\to\infty$. The latter limit can be understood by noting that
for large $A$ the electron density becomes flatter and flatter,
due to the strong Coulomb repulsion. Thus, in the limit
$A\to\infty$ we end up with a uniform electron density exactly
neutralized by the ion density background. For such a homogeneous
system, the Bohm-Gross dispersion relation holds, which for long
wavelengths yields $\Omega=\omega_0$. Indeed, if one computes the
average density using the prescription used for the EOS, one
obtains $\langle n\rangle = \int n^2 dx/\int ndx
=A/(\sqrt{2}\sigma_0) \to 1$, for $A\to\infty$ (see the inset of
Fig. \ref{fig2}). Thus, as expected, for large Coulomb effects the
average electron density becomes equal to the ion background
density.

{\it Simulations}.--- In order to check the validity of the above
results, we performed numerical simulations of the Wigner-Poisson
(WP) system, which is equivalent to the time-dependent Hartree
equations \cite{Manfredi3}. In the normalized variables, the
Wigner pseudo-probability distribution $f(x,v,t)$ satisfies the
evolution equation
\begin{equation}
\label{wig} \frac{\partial f}{\partial t} +  v\,\frac{\partial
f}{\partial x} -  i\int\!\frac{dx'\,dv'}{2\pi\,H^2}\, \delta
V_{\rm eff}\,e^{\frac{i(v-v')x'}{H}}f(x,v',t) = 0 ,
\end{equation}
and is coupled to the Poisson equation. In Eq. (\ref{wig}),
$\delta V_{\rm eff} \equiv V_{\rm eff}(x+x'/2,t) - V_{\rm
eff}(x-x'/2,t)$. It is important to note that this is a
microscopic quantum mean-field model, much more general than the
hydrodynamic model on which our Lagrangian theory was based.
\begin{figure}[htb]
\centering
\includegraphics[width=6.cm]{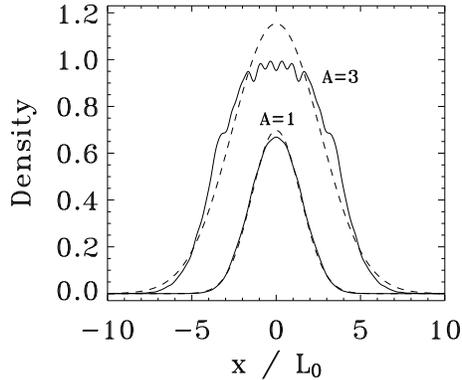}
\caption{Solid lines: Electron density profiles at $\omega_0t=150$
from the WP simulations for $H=0.5$ and two values of $A$. The
dashed lines represent Gaussian distributions with the same width
and the same area as the numerical curves.} \label{fig3}
\end{figure}

The initial condition used in the simulations is a quantum
canonical distribution for the harmonic oscillator at finite
temperature \cite{Imre}, where the spatial width $\sigma_0$ has
been adjusted to the value obtained from the Lagrangian approach
to account for the Coulomb repulsion. This is very close, but not
quite identical, to an exact equilibrium of the WP equations, so
that the width of the electron density starts to oscillate. We
then compute the evolution of the dispersion $\langle x^2
\rangle^{1/2} = (\int fx^2 dxdv/\int f dxdv)^{1/2}$ and its
frequency spectrum, which generally shows a sharp peak at a
dominant frequency.

The results of the WP simulations are plotted in Fig. \ref{fig2}
(squares) and agree very well with the theoretical curve based on
the Lagrangian approach. The agreement slightly deteriorates for
larger values of $A$, because the electron density deviates from
the Gaussian profile due to strong Coulomb repulsion. This is
clearly visible in Fig. \ref{fig3}, where we represent the evolved
density profiles for two values of $A$. For $A=1$, the profile is
still approximately Gaussian, whereas for $A=3$ an intricate
internal structure has developed. Nevertheless, even in this case,
the error on the frequency is still just over 3\%.
\begin{table}[htbp]
\caption{\label{table1} The breather frequency $\Omega$ and
the equilibrium width $\sigma_0$ for $A=1$ and for various values of $H$.}
\begin{ruledtabular}
\begin{tabular}{cccc}
$H$ & $\sigma_0$ & $\Omega$ (theory) & $\Omega$ (sim.)\\ \hline
0.00 & 1.41  & 1.58  & 1.60 \\
0.50 & 1.43  & 1.59  & 1.60 \\
1.00 & 1.47  & 1.60  & 1.51 \\
1.50 & 1.52  & 1.61  & 1.57 \\
2.00 & 1.59  & 1.63  & 1.63 \\
3.00 & 1.73  & 1.66  & 1.70 \\
\end{tabular}
\end{ruledtabular}
\end{table}

Table \ref{table1} shows that the breather frequency depends
weakly on the parameter $H$ (and hence on the electron
temperature). Our theoretical results are in good agreement with
the simulations, except for $H=1$. For this case, the frequency
spectrum is particularly broad, denoting a significant
fragmentation of the resonance.

{\it Nonparabolic wells}. --- For a parabolic potential well,
there is no coupling between the breather and dipole modes, which may
render the experimental detection of the breather mode by optical
means difficult to realize in practice.

It can be shown that nonparabolic corrections do not introduce any
linear coupling if the confining potential is symmetric. A more
interesting situation arises for {\em asymmetric} wells
\cite{asymm}, which we model by adding a small cubic term to the
confining potential, $V_{\rm cub}(x)=(K/3)x^3$. The equations of
motion are then
\begin{eqnarray}
\label{cub1} \ddot{d} &+& d = -K(\sigma^2 + d^2), \\
\label{cub2} \ddot{\sigma} &+& \sigma = \frac{\sqrt{2}\,A}{2} +
\frac{\sqrt{3}\,A^2}{3\,\overline{n}^2\,\sigma^3}  +
\frac{H^2}{4\,\sigma^3} -2K\sigma d \,.
\end{eqnarray}
Linearizing (10) and (11) around the stable fixed point $(d_0,\sigma_0)$,
we indeed find a coupling between the breather and the dipole, with
resonant frequencies $\Omega_\pm$.
\begin{figure}[htb]
\centering
\includegraphics[width=6.5cm]{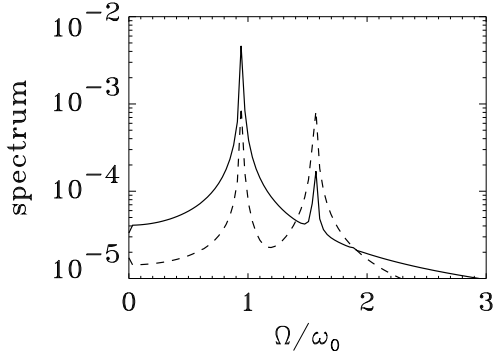}
\caption{The frequency spectra of the dipole (solid line) and
breather (dashed) modes, for an asymmetric well with $A=1$,
$H=0.5$, $K=-0.1$.} \label{fig4}
\end{figure}

We model the coupling to the laser field by instantaneously
shifting the initial dipole of a small quantity, i.e.,
$d(0)=d_0+\widetilde{d}$. Figure \ref{fig4} shows a typical
spectrum obtained from the numerical solution of Eqs.
(\ref{cub1})-(\ref{cub2}), which is proportional to the optical absorption spectrum
commonly measured in the experiments. The sharp peaks correspond to the
resonant frequencies $\Omega_-=0.97$ and $\Omega_+=1.58$, which
are rather close to those obtained for parabolic confinement.

The breather mode can thus be triggered using a purely dipolar
excitation, and a clear signature of the breather frequency can be
observed in the optical absorption spectrum. This opens the way to optically
detecting the breather mode by means of standard pump-probe
experiments. Finally, the methods used here could be readily extended
to 3D nanostructures, and may find applications in related areas such as
quantum free-electron lasers \cite{Piovella}.

This work was partially supported by the Alexander von Humboldt
Foundation and by the Agence Nationale de la Recherche.


\begin{thebibliography}{99}

\bibitem{Zoller} P. Zoller {\it et al.} Eur. Phys. J. D {\bf 36}, 203 (2005).

\bibitem{Heyman} J. N. Heyman {\it et al.}, Appl. Phys. Lett. {\bf 72}, 644 (1998).

\bibitem{Kohn} W. Kohn, Phys. Rev. {\bf 123}, 1242 (1961).

\bibitem{Harakesh} M. N. Harakesh {\it et al.}, Phys. Rev. Lett. {\bf 38}, 676 (1977).

\bibitem{Youngblood} D. H. Youngblood {\it et al.}, Phys. Rev. Lett. {\bf 39}, 1188 (1977).

\bibitem{Goral} K. G\'oral {\it et al.}, Phys. Rev. A {\bf 67}, 025601 (2003).

\bibitem{Zhou} Y. Zhou and G. Huang, Phys. Rev. A {\bf 75}, 023611 (2007).

\bibitem{Portales} H. Portales {\it et al.}, J. Chem. Phys. {\bf 115}, 3444 (2001).

\bibitem{Santer} M. Santer {\it et al.}, Phys. Rev. Lett. {\bf 89}, 286801 (2002).

\bibitem{Wijewardane} H. O. Wijewardane and C. A. Ullrich, Appl. Phys. Lett. {\bf 84}, 3984 (2004).

\bibitem{Manfredi} G. Manfredi and F. Haas, Phys. Rev. B {\bf 64}, 075316 (2001).

\bibitem{Manfredi2} G. Manfredi, Fields Inst. Commun. {\bf 46}, 263 (2005).

\bibitem{Crouseilles} N. Crouseilles {\it et al.}, Phys. Rev. B {\bf 78}, 155412 (2008).

\bibitem{Gardner} C. L. Gardner and C. Ringhofer, Phys. Rev. E {\bf 53}, 157 (1996).

\bibitem{Manfredi3} G. Manfredi and P.-A. Hervieux, Appl. Phys. Lett. {\bf 91}, 061108 (2007).

\bibitem{Gusev} G. M. Gusev {\it et al.}, Phys. Rev. B {\bf 65}, 205316 (2002).

\bibitem{Heyman95} J. N. Heyman {\it et al.}, Phys. Rev. Lett. {\bf 74}, 2682 (1995)

\bibitem{Matuszewski} M. Matuszewski {\it et al.}, Phys. Rev. Lett. {\bf 95}, 050403 (2005).

\bibitem{Dobson} J. F. Dobson, Phys. Rev. Lett. {\bf 73}, 2244 (1994).

\bibitem{Imre} K. Imre {\it et al.}, J. Math. Phys. {\bf 8}, 1097 (1967).

\bibitem{asymm} E. Rosencher and P. Bois, Phys. Rev. B {\bf 44}, 11315 (1991); E.
Dupont {\it at al.}, IEEE J. Quantum Electron. {\bf 42}, 1157 (2006).

\bibitem{Piovella} N. Piovella {\it et al.}, Phys. Rev. Lett. {\bf 100}, 044801 (2008).

\end{thebibliography}
\end{document}